\documentclass[12pt]{article}

\usepackage{graphicx}
\usepackage{amsmath,amsthm}
\usepackage{amssymb}
\usepackage{graphicx}
\usepackage{xcolor}
\usepackage{ragged2e}
\usepackage{svg}

\definecolor{ceruleanblue}{rgb}{0.16, 0.32, 0.75}
\definecolor{cardinal}{rgb}{0.77, 0.12, 0.23}
\definecolor{grey}{rgb}{0.3, 0.3, 0.3}
\definecolor{orange2}{RGB}{225,147,0}

\usepackage[colorlinks=true, linkcolor=blue, citecolor=blue, urlcolor=blue]{hyperref}
\usepackage{pgf,tikz}
\usepackage[top=2.5truecm,bottom=2.5truecm,left=2.5truecm,right=2.5truecm]{geometry}

\usepackage{authblk}

\usepackage{lscape}
\usepackage{longtable}
\usepackage{rotate}
\usepackage{rotating}
\usepackage{comment}
\usepackage[labelfont=bf,textfont=md]{caption}
\usepackage{subcaption}
\usepackage{float}
\usepackage{soul}
\usepackage{bigdelim}
\usepackage{setspace}

\usepackage{makecell}
\usepackage{booktabs}
\usepackage[toc,page]{appendix}

\usepackage{array,multirow,graphicx}

\usepackage{threeparttable}

\usepackage{tikz}

\definecolor{webgreen}{rgb}{0,.5,0}
\definecolor{webbrown}{rgb}{.6,0,0}
\definecolor{webblue}{rgb}{0,0,.7}
\definecolor{webyellow}{rgb}{0.98,0.92,0.73}

\usepackage[round]{natbib}
\bibpunct[, ]{(}{)}{;}{a}{,}{,}

\usepackage{svg}
\usepackage{url}

\usepackage{xcolor}
\usepackage{hyperref}

\hypersetup{
    colorlinks=true,
    linkcolor=black,      
    citecolor=black,      
    urlcolor=blue         
}

\usepackage{xpatch}
\xpatchbibmacro{title}{%
  \printfield{title}%
}{%
  \iffieldundef{url}{%
    \printfield{title}%
  }{%
    \href{\thefield{url}}{\printfield{title}}%
  }%
}{}{}

\begin{document}

\begin{singlespacing}
    
\title{\Large{\textsc{Artificial Intelligence and the US Economy: An Accounting Perspective on Investment and Production\thanks{
The views expressed in this paper do not necessarily reflect those of the Bank of Italy or ESCB. The authors would like to thank Francesco Paolo Conteduca for sharing and support on the Trade Monitor Data. They also thank Riccardo Cristadoro for comments.}}}}

\author{Luisa Carpinelli\thanks{Bank of Italy, Economics, Statistics and Research Department; \href{mailto:luisa.carpinelli@bancaditalia.it}{luisa.carpinelli@bancaditalia.it}}\qquad Filippo Natoli\thanks{Bank of Italy, Economics, Statistics and Research Department; \href{mailto:filippo.natoli@bancaditalia.it}{filippo.natoli@bancaditalia.it}}\qquad Marco Taboga\thanks{Bank of Italy, Economics, Statistics and Research Department; \href{mailto:marco.taboga@bancaditalia.it}{marco.taboga@bancaditalia.it}} \newline }

\date{\today}

\graphicspath{{graphs/}}

\maketitle
\thispagestyle{empty}

\begin{abstract}
    \singlespacing
Artificial intelligence (AI) has moved to the center of policy, market, and academic debates, but its macroeconomic footprint is still only partly understood. This paper provides an overview on how the current AI wave is captured in US national accounts, combining a simple macro-accounting framework with a stylized description of the AI production process. We highlight the crucial role played by data centers, which constitute the backbone of the AI ecosystem and have attracted formidable investment in 2025, as they are indispensable for meeting the rapidly increasing worldwide demand for AI services. We document that the boom in IT and AI-related capital expenditure in the first three quarters of the year has given an outsized boost to aggregate demand, while its contribution to GDP growth is smaller once the high import content of AI hardware is netted out. Furthermore, simple calculations suggest that, at current utilization rates and pricing, the production of services originating in new AI data centers could contribute to GDP over the turn of the next quarters on a scale comparable to that of investment spending to date. Short reinvestment cycles and uncertainty about future AI demand -- while not currently acting as a macroeconomic drag -- can nevertheless fuel macroeconomic risks over the medium term.

\end{abstract}

\end{singlespacing}

\bigskip

\noindent \qquad \quad \, \footnotesize{\textit{ JEL:} E2, O3, O4}

\noindent \qquad \qquad \footnotesize{\textit{Keywords:} artificial intelligence, capital expenditures, data centers, national accounts} \\
\newpage

\section{Introduction}\label{sec:introduction}
\normalsize

The rapid development and adoption of artificial intelligence (AI) technologies have been at the forefront of policy, market, and academic debates. From a financial stability perspective, a recurring question is whether the sharp appreciation in the share prices of technology companies is justified by a credible and sustainable expansion in their underlying profitability. From a more structural economic perspective, the key theme is how AI adoption will affect productivity and employment.

\begin{figure}[H]
\begin{center}
\caption{ \textsc{Private construction spending by selected industries.} }
\vspace{-1mm}
\includegraphics[width=0.9\textwidth]{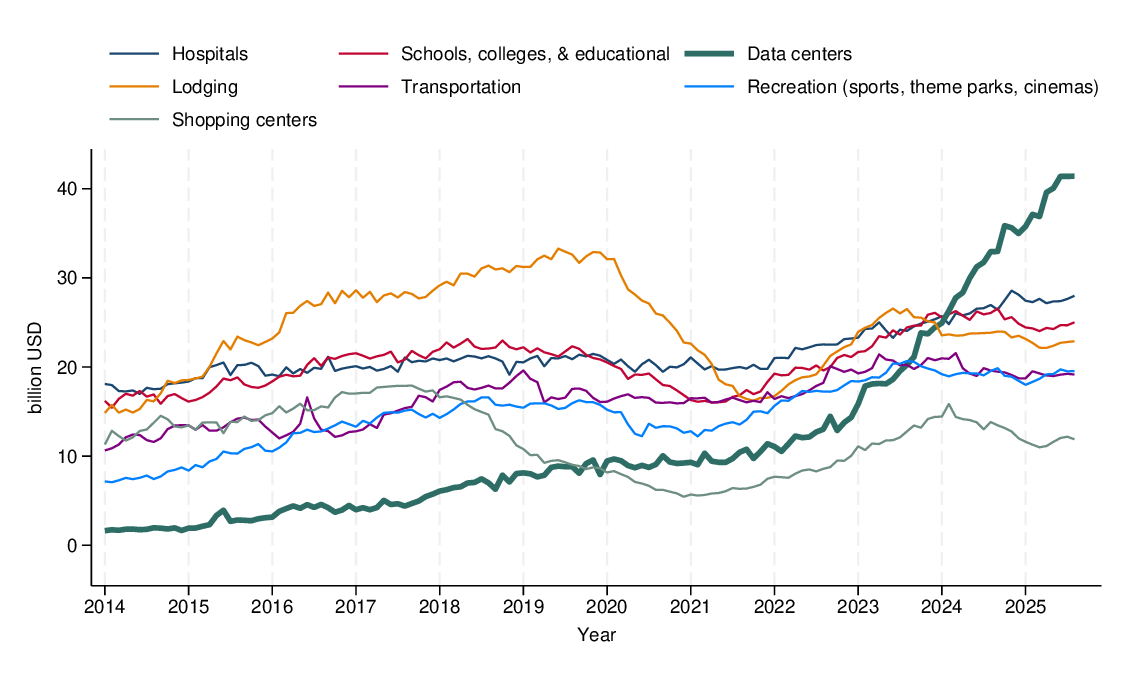}
\label{fig:construction}
\vspace{-2mm}
\footnotesize{\textit{\\Source. U.S. Census Bureau, Value of Construction Put in Place. Last observation: August 2025.}} 
\end{center}
\end{figure}

More recently, a conjunctural issue attracting widespread attention from the media and market analysts is the extent to which AI‑related capital investment is currently contributing to US GDP dynamics. US growth in 2025 has been resilient above expectations, and some narratives suggest that investments primarily related to the rapid acceleration in the build‑up of data centers (Figure \ref{fig:construction}) have been the engine of this resilience.

This paper provides a tentative
assessment of how AI production is being reflected in GDP components. 
We combine a macro-accounting perspective with a detailed description of the AI production ecosystem to answer a simple question: how, and through which channels, is the current AI wave showing up in measured US GDP? We focus on the parts of the AI value chain that are already large enough to matter for the short-term outlook – notably investment in data‑center infrastructure and the provision of cloud and AI services – and on the distinction between domestic value added and foreign content embedded in AI‑related spending.

Assessing the incidence and channels of AI for the economic outlook is highly relevant for several reasons.
For starters, it can guide policymaking, in particular monetary policy, by improving the understanding of inflationary pressures and labor market dynamics -- for instance, by helping to reconcile the apparent dichotomy between resilient overall growth and a weak labor market.
More broadly, a clear mapping of how AI investment and production enter GDP is essential for evaluating the strength and composition of demand, for gauging the domestic versus foreign share of value added in AI supply chains, and for informing fiscal and industrial policies aimed at supporting high-value activities. It also matters for financial stability analysis, as it helps distinguish between genuine, income-backed growth in AI-related sectors and developments that are mainly driven by asset price revaluations.

Quantification is nevertheless fraught with methodological and measurement challenges. First, isolating AI-related activity in the national accounts is difficult because AI is typically embedded in broader aggregates -- such as IT equipment, semiconductors, software, R\&D, and IT services -- and because it is often impossible to pin down the services where AI is incorporated. In practice, statistics frequently allow to capture only a broad ``technology'' sector rather than AI specifically. Second, firm-level AI figures are also rarely observable: key players in the AI production ecosystem are large, multi-activity firms whose financial statements combine costs, revenues, and profits across many lines of business. As a result, isolating the contribution of AI to costs, revenues and profits is extremely challenging. Third, the broader “second-order” effects set in motion by the AI transformation, including changes in productivity and employment linked to its adoption, are still too nascent to be reliably assessed.  As such, they are  outside the scope of this paper, which adopts a mechanical accounting perspective on how AI surfaces in GDP components.

Three findings stand out. First, the boom in IT and AI‑related capital expenditure has provided exceptional support to aggregate demand, but its net contribution to US GDP growth is more limited than headline investment figures suggest because a large share of the associated hardware is imported. This reflects the fact that, while activities like chip design, data center operation, cloud services, and model development are mainly domestic,  only part of the value created along the AI hardware supply chain accrues to the United States, as the fabrication and assembly of servers and accelerators are concentrated abroad. 

Second, data centers are clearly the backbone of the AI ecosystem. They concentrate the massive computational capacity required for AI model training, inference, and cloud‑based services, and translate AI‑related capital spending into productive output. From a macroeconomic standpoint, the construction and equipment procurement of data center facilities generate significant demand for construction services, high‑performance hardware and ancillary inputs, while the computational services they provide appear in the national accounts as final consumption and constitute intermediate inputs for other industries. Hence, data center activity is a pivotal channel through which AI investment feeds aggregate demand and contributes to GDP growth, making it a central focus of any macroeconomic analysis of the AI wave.

Third, as the new generation of AI data centers becomes fully operational and sells computational and AI services to final users at home and abroad, the resulting revenue stream is expected to add to GDP at a level comparable to the capital expenditure needed to expand the underlying AI production capacity. This follows from the high utilisation rates of existing data center infrastructure, the current pricing of GPU‑based services, and the short pay‑back periods estimated in the paper. 

The rest of the paper is organized as follows.
Section~\ref{sec:roadmap} provides a roadmap to the AI industry, describing the main actors in the ecosystem -- hardware vendors, providers of cloud infrastructure and AI labs -- and highlighting the central role of data centers and global supply chains.
Section~\ref{sec:investment} turns to AI investment and quantifies how much of the strength in US demand and output in the first half of 2025 can be attributed to technology‑related capital expenditures once imports and sectoral value added are taken into account.
Section~\ref{sec:production} describes the production of AI services using the example of an AI data center to illustrate the channels through which computational services enter national accounts.
Sections~\ref{sec:future} and~\ref{sec:conclusions} discuss future developments and draw conclusions.

\section{A roadmap to the AI industry}
\label{sec:roadmap}

In this section we provide a stylized description of how the AI industry is structured, which helps to understand the channels through which AI activity affects the economy. 

The AI ecosystem is organized around three often overlapping groups of actors: 1) hardware vendors, who design, build and sell the computer equipment used in AI data centers; 2) cloud infrastructure providers, who own the data centers and rent out the computing capacity; 3) providers of AI software, who demand computing capacity to run AI models and sell the derived services to their customers.

We describe these groups in the following subsections, paying particular attention to where the main firms are headquartered and where different stages of production take place. This geographical dimension is crucial for understanding how much of the value created along the AI supply chain is recorded as domestic output in the United States and how much accrues to trading partners.

\subsection{Hardware vendors}

Modern AI workloads are computationally intensive and need to be run on specialized hardware (so-called AI chips), such as Graphical Processing Units (GPUs) and Tensor Processing Units (TPUs). By leveraging thousands of computational cores working in parallel, these chips accelerate the mathematical operations, such as matrix multiplication, that recur most often in AI-model training and inference. The acceleration factor is several orders of magnitude with respect to classical CPU (Central Processing Units) chips. 

The production of AI chips -- the most costly item in AI data centers -- and their packaging into boards and server racks happens along a complex global supply network:

\begin{itemize}

\item Chip design: the design of AI chips is mostly performed by large US public companies such as NVIDIA (the leading player in this space), AMD and Broadcom (which co-designs proprietary chips with large data-center operators like Apple and Google).

\item Chip manufacturing: once chips are designed, they are manufactured in semiconductor fabrication plants (so-called fabs). These are mostly operated by Asian firms such as TSMC (Taiwan) and Samsung (Korea), but some US plants have begun production on the back of public financial incentives provided by the Chips Act. Memory modules, expensive components tightly integrated in AI chips, are provided both by US manufactures (Micron) and Asian ones.

\item Server assembly: packaged GPU modules go to server makers (e.g., SuperMicro, Hewlett Packard, Dell), who integrate them into full servers and racks comprising thousands of other less costly components (e.g., networking modules, mechanical, electrical and cooling parts).\footnote{The manufacturers either sell under their own brand (\textit{Original Equipment Manufacturers}, OEM), or build as custom for someone else’s brand (\textit{Original Design Manufacturers}, ODM). Big-name server OEMs are Dell Technologies (PowerEdge), HPE (ProLiant, Synergy), Lenovo (ThinkSystem); major server ODMs are Quanta Computer (and QCT – Quanta Cloud Technology), Wiwynn (Wistron spinoff focused on cloud DC), Inventec, Foxconn / Hon Hai. } Given the high number of components involved, this segment of the value chain is much more difficult to trace, although it is known to be dislocated not only in Asia and the US, but also in other countries that are geographically close to data-center locations (e.g., Mexico).
\end{itemize}

This already complex and geographically spread supply chain is complemented by numerous other US and foreign companies both upstream (providers of chip manufacturing equipment; e.g., ASML, Applied Materials) and downstream (integrators and providers of data-center equipment; e.g. Schneider Electric). 

From an economic perspective, the main segments of the hardware industry described above function as oligopolies. They are dominated by a small number of firms that enjoy high profit margins and are shielded by substantial barriers to entry. This is especially evident in chip manufacturing, where producing cutting-edge chips requires fabrication plants that not only cost tens of billions of dollars to build, but also demand highly specialized, hard-to-replicate know-how and technical expertise to fine-tune production processes and achieve economically viable yield levels.

\subsection{Cloud infrastructure providers}

Cloud infrastructure providers own and operate globally distributed data centers. They lease the computing capacity of these data centers to organizations who demand computing, file storage, networking and related services (Infrastructure as a Service - IaaS). 

The largest providers -- often referred to as hyperscalers -- are Amazon Web Services (AWS), Microsoft Azure, and Google Cloud Platform (GCP). Together, they account for 65\% of the global cloud-computing market \citep{omdia2025cloudspending}. Meta is sometimes classified as a “quasi-hyperscaler” because its global data-center footprint is comparable, although its infrastructure primarily supports its own applications. Other, smaller providers include Oracle Cloud, IBM Cloud, and CoreWeave.

These firms supply cloud infrastructure both for conventional services (e.g.  web hosting and database management) and AI workloads such as model training and inference. Because hyperscalers do not provide a formal revenue and profit split between AI‑related and traditional cloud services, it is difficult to isolate AI activity, especially since conventional services likely still represent the majority of their cloud revenues.

A common pricing model of cloud-infrastructure services is pay-as-you-go, whereby AI compute is billed per GPU-hour (or per second/minute), with separate charges for storage and network usage. These hourly prices are published by hyperscalers on their web sites in real time, which facilitates the assessment of demand and supply conditions in the market. However, pricing conditions are not always common knowledge, as, in addition to on-demand usage, services are often delivered through larger, negotiated contracts with enterprises. Such agreements typically involve committed spend (e.g., a minimum annual spend) and include enterprise-grade features like enhanced security, compliance assurances, administrative controls, data-handling guarantees, dedicated capacity, and dedicated support. 

Cloud infrastructure providers incur a range of costs. These include (i) developing and maintaining the software layer that enables resource leasing, (ii) ensuring security, monitoring performance, and maintaining reliability, (iii) covering electricity and other operating expenses, and (iv) most critically, amortizing the substantial physical investment in data centres, which includes not only computing hardware but also land, buildings, and ancillary systems such as power, cooling, and networking. The cost associated to physical investments has become more and more relevant as AI-related services have gained traction. Although AI-related cloud services can appear similar to the traditional cloud services, the underlying infrastructure differs substantially. Traditional workloads run on CPU‑centric server racks, whereas AI workloads rely on clusters built around large arrays of costly AI accelerators (GPUs and TPUs). These AI‑focused facilities require much higher power density, advanced cooling, high‑speed networking, and high‑throughput storage. All of these factors drive up the costs of AI infrastructure substantially.

From a geographical standpoint, all of the players mentioned above (Amazon, CoreWeave, Google, IBM, Microsoft, and Oracle) are headquartered in the United States and build data centers worldwide. While the precise distribution of their data centers across countries is not publicly known, most estimates suggest that roughly 50\% of global data center capacity is located in the US.\footnote{See, e.g., \citet{synergy2025hyperscale}, \citet{neufeld2025datacenter}, \citet{ciolli2025america}, \citet{leichter2025datacenters}.} A substantial share of the remaining capacity is in China, where several large domestic providers operate, most notably Alibaba Cloud, followed by Huawei Cloud, Tencent Cloud, and Baidu Cloud.  However, it is unclear how this geographic distribution translates into the revenues and profits reported by US-domiciled subsidiaries, or how these figures are influenced by cross-border intra-group transfer pricing.

\subsection{AI labs}

AI labs are companies that train frontier AI models and sell services derived from these models. Some of the main labs (and their GenAI models) are OpenAI (ChatGPT), Anthropic (Claude), Google DeepMind (Gemini, Gemma), Meta AI (Llama, Meta AI assistant), xAI (Grok), Mistral, DeepSeek, Z.AI (GLM models), as well as the AI divisions of Chinese tech giants Alibaba (Qwen), Baidu (ERNIE) and Tencent (Hunyuan). 

AI labs' revenues come mainly from selling the following two services:
\begin{itemize}
\item AI as a Service (AIaaS): AI laboratories expose their models through Application Programming Interfaces (APIs), enabling organizations to submit queries programmatically and receive model‑generated responses. This kind of service is typically demanded by corporations that develop and sell AI-based or AI-enhanced software, or by firms that use AI to automate their workflows. For Large Language Models, the service is typically priced on a per‑million‑token basis (one token roughly corresponds to a single English word generated by a model). 
\item Software-as-a-Service (SaaS): AI models are integrated into user‑facing applications that are sold on a per‑seat basis through monthly or annual subscriptions, both to enterprises and to consumers. OpenAI’s GPT‑based and Google's Gemini chatbots, as well as Microsoft's Copilot products are some examples of popular SaaS offerings.
\end{itemize}

Another less common source of revenue are licensing contracts, such as those in place between OpenAI and Microsoft, which allow the latter to directly use OpenAI models in its services.

The largest costs of AI labs are associated with recruiting and retaining top talent and with procuring the enormous computational resources that are needed to train and serve models. Several labs (e.g., OpenAI and Anthropic) do not own large data centers and satisfy their computing needs by purchasing cloud infrastructure services (see previous sub-section).

The segment of AI labs remains particularly opaque. 
Many labs are often large vertically integrated conglomerates, such as Google and Meta, that do not disclose the revenues and profits of their AI research units. Other labs, dedicated almost entirely to model development (e.g., Anthropic, OpenAI), are privately held and therefore their financial performance is not publicly available. Moreover, there are often partnerships and cross-holdings between the two groups of labs/companies.\footnote{A few examples. Between 2019 and 2023 Microsoft invested over \$13 billion in OpenAI, securing a claim on roughly 49\% of OpenAI LP’s future profits and integrating OpenAI’s models deeply into Azure for cloud computing and Microsoft 365. Amazon and, with a lower investment, Google are partners/investors in Anthropic, that runs its models largely on AWS as its primary cloud provider.} 

From a geographic standpoint, the largest labs are controlled by companies headquartered in the US (Anthropic, Google DeepMind, Meta AI, OpenAI, xAI) and China (Alibaba, Baidu, DeepSeek, Tencent, Z.AI), with the notable exception of Mistral (France).

\subsection{General remarks on the AI ecosystem}

Although simplified, the above depiction of the AI ecosystem highlights several features.
First, the supply chain for the manufacturing and assembly of data-center hardware is truly global, while cloud infrastructure provision and AI services are more domestically concentrated.

Second, data centers form the backbone of the ecosystem, absorbing the bulk of capital expenditure and enabling almost all AI-related revenue streams: for cloud infrastructure providers, the data center is the product; for model labs, data centers are probably the largest input cost; for chip and server manufacturers, data centers are the primary deployment location for their products and thus the main source of demand. In this sense, AI production is essentially a data‑center-centric industrial system.

Third, in such an articulated and complex ecosystem, AI-related activity affects GDP through several measurable channels. Data-center building expenses and AI-related R\&D (for example, related to chip design and model development)  sustain aggregate investment. Chip and server manufacturing significantly impacts import and export, given its global dislocation. The production of AI services affects the magnitude of value added and its distribution across industries. Use of AI services by final users impacts consumption and public expenditures. 
The next sections provide quantitative evidence about these effects.

\section{AI Investment and Aggregate Demand}
\label{sec:investment}

\begin{figure}[H]
\begin{center}
\caption{ \textsc{Growth in technology investments} }
\vspace{-15mm}
\includegraphics[width=1\textwidth]{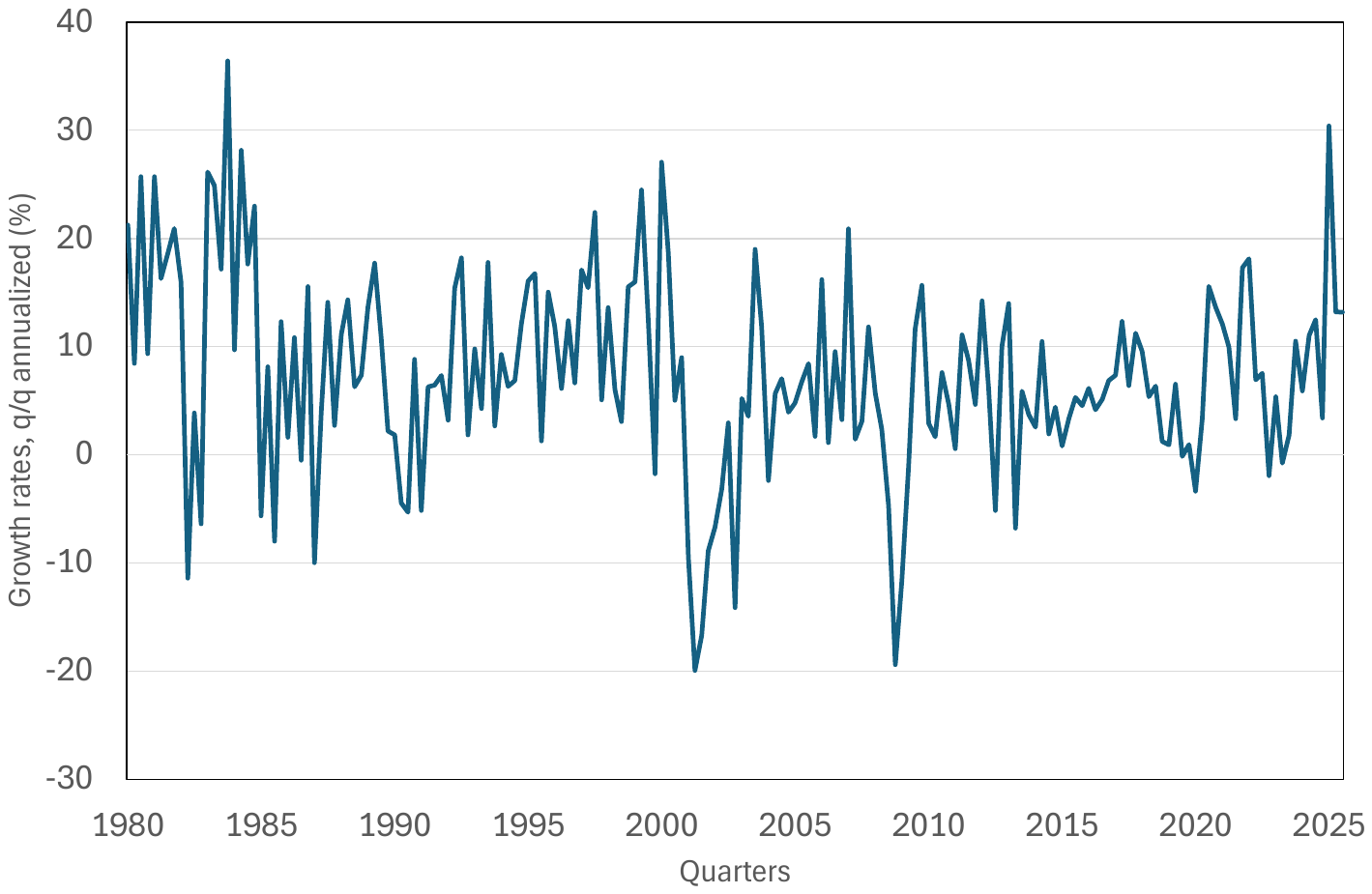}
\vspace{-20mm}
\footnotesize{\textit{\\Note: Quarter-on-quarter annualized growth in nominal investment in information processing equipment and software. Last observation: 2025 Q3. Source: U.S. Bureau of Economic Analysis.}} 
\label{fig:invlr}
\end{center}
\end{figure}

In the first half of 2025, real private fixed investment surged in the US (7.1\% in Q1 and 4.4\% in Q2), followed by a more moderate expansion in Q3 (1\%). Aggregate investment dynamics reflected that of IT‑related capital expenditure,\footnote{This refers to the category labelled private \emph{Fixed investment in information processing, equipment and software} in the Bureau of Economic Analysis statistics. It is an ad hoc measure of IT‑related investment, summing three different subitems of the investment aggregate: (1) \emph{Computer and peripheral equipment}; (2) \emph{Other} (other related equipment from the category of equipment investment); (3) \emph{Software} (from the intellectual property products category).} which grew by 36\%, 20\%, and 5\% (measured as annualized quarterly growth over the previous quarters) in the three quarters, respectively. These are record‑high numbers; an expansion like that of Q1 was unseen in nominal terms since the launch of IBM's personal computers in the 1980s (Figure \ref{fig:invlr}). 

\begin{figure}[H]
\begin{center}
\caption{ \textsc{Investment growth and contribution of tech vs non-tech items.} }
\vspace{-1cm}
\includegraphics[width=1\textwidth]{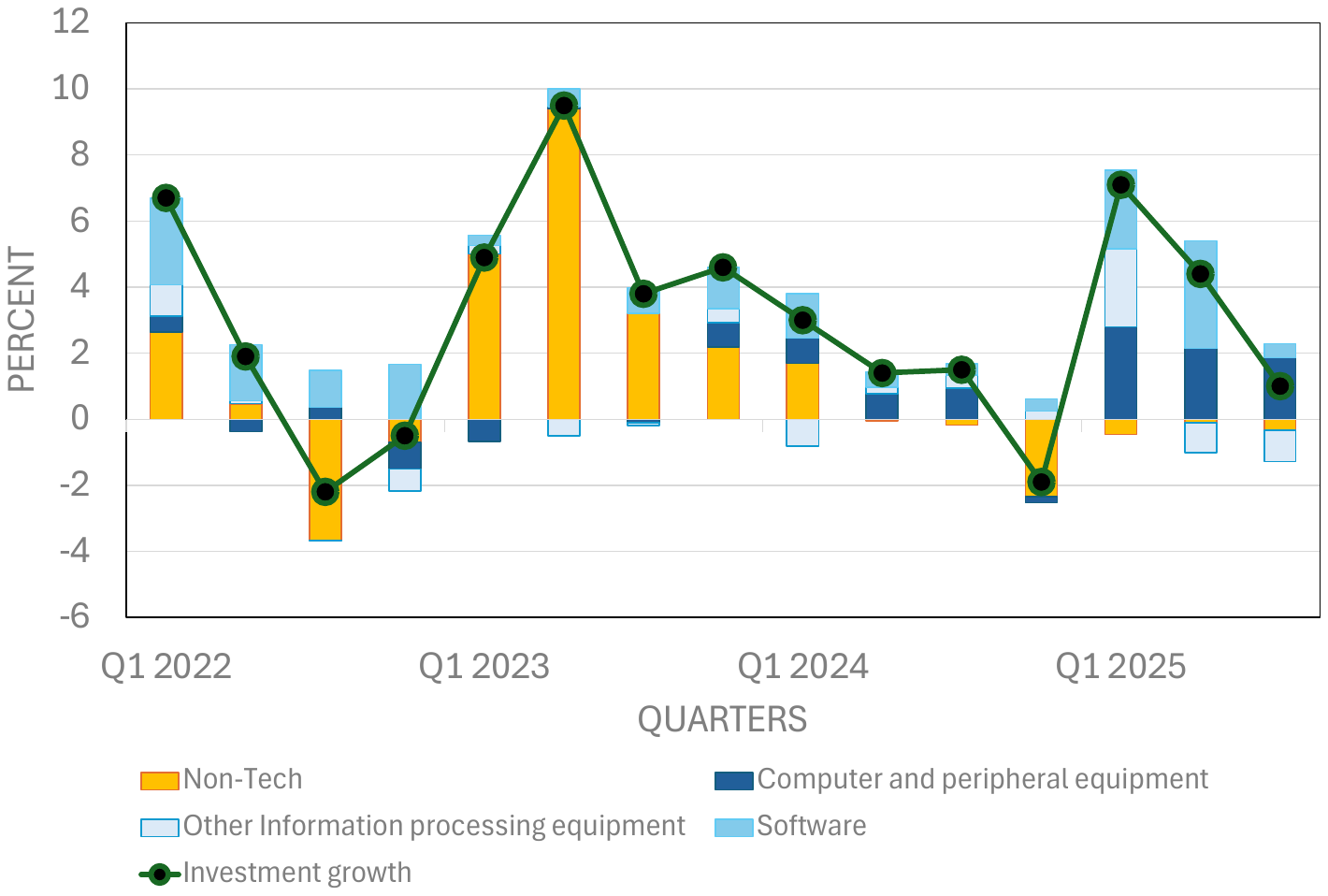}
\vspace{-1.5cm}
\raggedright
\footnotesize{\textit{\\Note: Black dotted line indicates quarter-on-quarter annualized growth (percent); colored bars indicate contribution to growth (percentage points). Last observation: 2025 Q3. Source: U.S. Bureau of Economic Analysis.}} 
\label{fig:inv}
\end{center}
\end{figure}

IT‑related capital expenditure provided indeed the largest positive contribution to aggregate investment expansion (7.6, 4.5 and 1.3 percentage points, respectively), offsetting declines both in the residential and non‑residential structure categories. Investment in hardware (\textit{Computers and peripheral equipment}) in particular increased massively, with purchases of computers and related equipment up 104\%, 62\% and 46\% (annualised rates, see Table \ref{tab:inv} and Figure \ref{fig:inv}).\footnote{Software investment expansion, that contracted in Q3, was also less buoyant in the first half of the year, yet -- since its level is roughly three times that of hardware -- it has made a comparable contribution.}

\begin{table}[H]
\centering
\begin{minipage}{\linewidth}
\caption{Technology-related investments in National Accounts.}
\label{tab:inv}
\end{minipage}
\footnotesize
\scalebox{.85}{
\begin{tabular}{lcccccc}
\toprule
 & \multicolumn{2}{c}{2025 Q1} & \multicolumn{2}{c}{2025 Q2} & \multicolumn{2}{c}{2025 Q3} \\
\cmidrule(r){2-3} \cmidrule(r){4-5} \cmidrule(r){6-7}
 & q/q growth & contr. to I & q/q growth & contr. to I & q/q growth & contr. to I \\
\midrule
Private fixed investment (I) & 7.1 &  & 4.4 &  & 1.0 &  \\
\addlinespace
\quad Nonresidential & & & & & & \\
\quad\quad Equipment & & & & & & \\
\quad\quad\quad \textit{Computers and peripheral equipment} & 103.7 & 2.8 & 61.7 & 2.2 & 46.1 & 1.9 \\
\quad\quad\quad \textit{Other} & 42.5 & 2.4 & -12.7 & -0.9 & -13.7 & -1.0 \\
\addlinespace
\quad\quad Intellectual property products & & & & & & \\
\quad\quad\quad \textit{Software} & 18.8 & 2.4 & 26.6 & 3.3 & 3.1 & 0.4 \\
\addlinespace
\midrule
\quad\quad\textbf{Info. processing equip. and software} & 36.1 & 7.6 & 19.7 & 4.5 & 5.4 & 1.3 \\
\bottomrule
\end{tabular}}
\begin{minipage}{\linewidth}
\raggedright
\footnotesize
\vspace{.5cm}\textit{Note: Quarter-on-quarter growth rates are annualized and computed on real investment values. Contributions to quarterly annualized investment growth are expressed in percentage points. Source: U.S. Bureau of Economic Analysis.}
\end{minipage}
\end{table}

\begin{figure}[H]
\begin{center}
\caption{ \textsc{Capital expenditures of large tech companies.} }
\vspace{-1.5cm}
\includegraphics[width=1\textwidth]{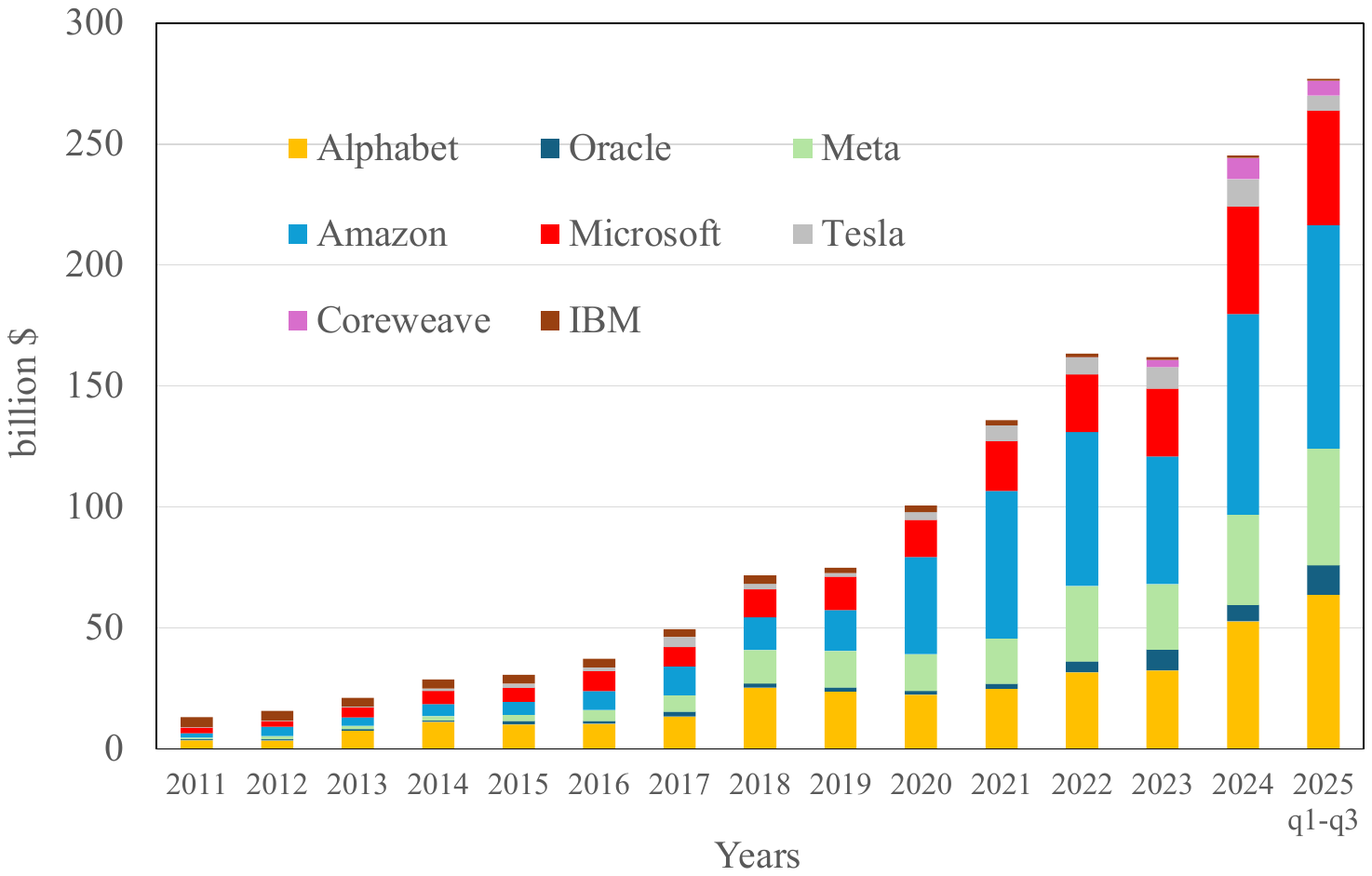}
\vspace{-1.5cm}
\footnotesize{\textit{\\Source: S\&P Capital IQ. Capital expenditures up to 2024 are computed for fiscal years.}} 
\label{fig:capex}
\end{center}
\end{figure}

The soaring investment is connected to the acceleration of capital expenditures in digital infrastructure for AI. The  component \textit{Computers and related equipment} also includes the purchase of servers (which embed AI chips such as NVIDIA's GPUs),  the costliest item  related to the construction of data centers – other expenses being building construction and the installation of networking, cooling and power equipment. Large tech companies made capital expenditures (capex henceforth) for \$245 billion overall in 2024, and more than \$275 billion in the first three quarters of 2025 (Figure \ref{fig:capex}).\footnote{Here we include both the largest firms selling cloud infrastructure/platform services to other organizations (Alphabet, Amazon, Microsoft, Oracle), two companies that are heavily investing in data centers but mostly for serving their own computational needs (Meta, Tesla), and also Corewave and IBM that have also developed data centers.}

Most of the forecasts now place overall AI-driven capex by large tech companies around \$300-350 billion for 2025. Companies' reports reveal indeed that a large chunk of this is related to the construction of data centers: for Microsoft, for example, nearly 100\% of the quoted 2025 capital expenditure of around US\$80 billion is explicitly for AI/data-center infrastructure; for Alphabet, ``the majority'' of the US\$75 billion is for servers/data centers; for Amazon ``the vast majority'' of over US\$100 billion is directed at ``data-center/AI infrastructure''.\footnote{Microsoft vice chair and President Brad Smith wrote in a blog post that in fiscal 2025 – which runs to 30 June – his company "is on track to invest approximately \$80 billion to build out AI-enabled data centres to train AI models and deploy AI and cloud-based applications around the world. More than half of this total investment will be in the United States”.  On a company’s earnings call, Alphabet CFO Anat Ashkenazi said a majority of that spending would “go towards our technical infrastructure, which includes servers and data centers”. Amazon CEO Andy Jassy said on a company’s earnings call referred that they expected to spend over \$100 billion in capex for the full year, the “vast majority” of which would go toward AI.}

This surge in investment prompted many economists, policymakers and commentators to deem AI investment as the prime engine behind US economic growth, if not the only one. Harvard economist Jason Furman said that in the first two quarters of 2025, 92\% of the increase in demand in the US economy was due to just two categories in GDP -- information processing equipment and software. Nobel Prize Paul Krugman seemed to agree that the ``AI boom is driving most, possibly all, of the economy's recent growth''. J.P.\ Morgan said that AI-related capital expenditures contributed 1.1 percentage points to GDP growth ``outpacing the US consumer as an engine of expansion''. A Deutsche Bank newsletter at the end of September read ``AI machines -- in quite a literal sense -- appear to be saving the US economy right now''.\footnote{See \cite{furman, NYT, jpm}.}

\begin{figure}[H]
\centering
\caption{\textsc{Investment and import of tech-related products}}
\label{fig:invimp2}

\begin{minipage}{0.5\textwidth}
    \centering
    \includegraphics[width=\linewidth]{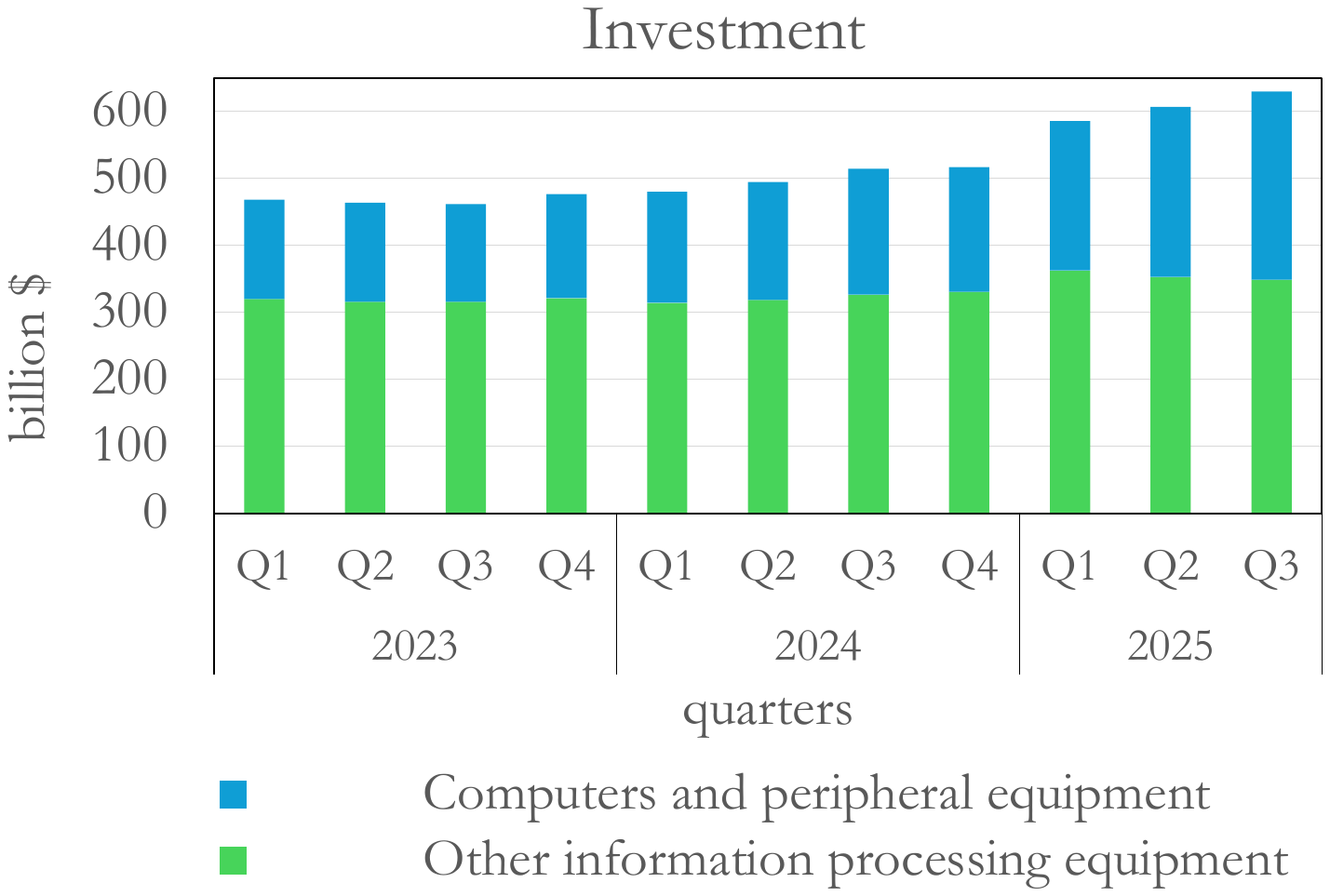}
\end{minipage}%
\begin{minipage}{0.5\textwidth}
    \centering
    \includegraphics[width=\linewidth]{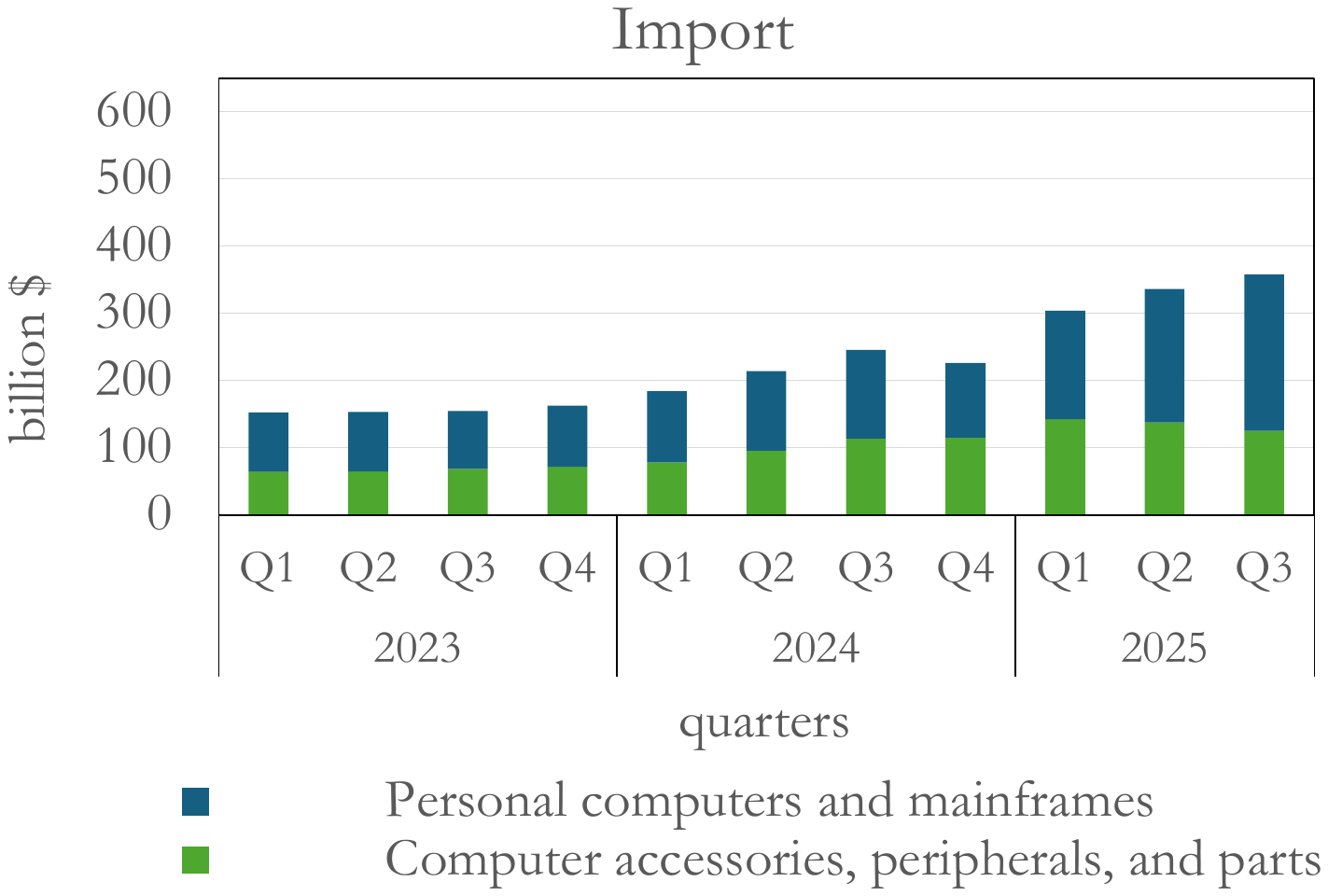}
\end{minipage}

\vspace{2mm}
\begin{minipage}{\linewidth}
\raggedright
\footnotesize
\textit{Note: Investment and imports of tech-related products, billions of dollars. Source: Bureau of Economic Analysis.}
\end{minipage}
\end{figure}

However, an analysis of national accounts reveals that, while AI investment strongly sustained demand, its impact on GDP growth was more limited, given that a substantial share of AI-related investment consisted of imported capital goods. In terms of macroeconomic accounting, imports are a drag on GDP dynamics for the US, while contributing to the GDP of the trading partner.\footnote{The role of imports is also acknowledged by \citet{gser, gs} and \citet{everc}.} The flow of imports of \textit{Computer, peripherals and parts} is more than half the value of total investment in \textit{Information processing equipment}. The surge of investments was indeed mirrored by the surge of imported computers and peripherals (+236\%, +47\%, and  +24\% in the first three quarters of the year, respectively, see Figure \ref{fig:invimp2}). In particular, the category of \textit{Personal Computers and Mainframes} is the one that includes servers. It should not be a surprise that this infrastructure is largely imported into the U.S., despite the main GPU producer -- NVIDIA -- being a U.S. company: as described in Section \ref{sec:roadmap}, both (i) the GPU chip manufacturing/packaging and (ii) a lot of the server integration happens outside the U.S..


\begin{figure}[H]
\begin{center}
\caption{ \textsc{Import of servers by country of origin} }
\vspace{-12mm}
\includegraphics[width=1\textwidth]{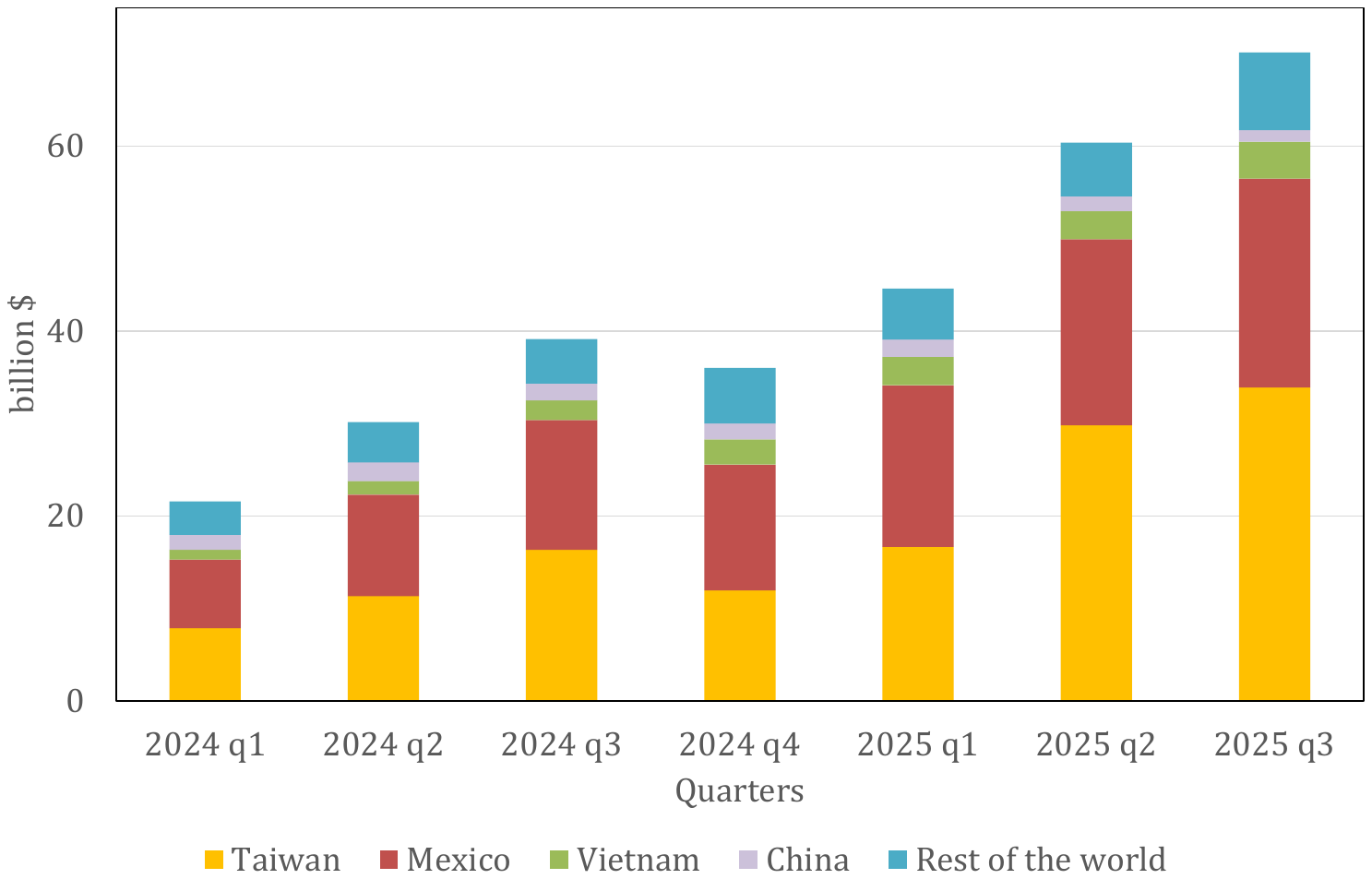}
\vspace{-15mm}
\footnotesize{\textit{\\Note: Billion dollars. Source: Trade Data Monitor.}} 
\label{fig:tdm}
\end{center}
\end{figure}

The relevant role played by the imports of servers is confirmed by Trade Monitor data, providing a more granular breakdown of trade flows. Figure \ref{fig:tdm} shows a proxy of this category of imports: their inflow in the first three quarters of 2025 was almost double the nominal value imported in the same 9 months in 2024. Imports came largely from Mexico and Vietnam. In particular, Mexico's growing role in the exports of AI-related products stands out, with this category now rivaling its traditionally key auto parts outflows, as also highlighted by specialized analysts \citep{CEmexexp, CEmex}

\begin{figure}[H]
\begin{center}
\caption{ \textsc{Contribution to growth, tech vs non-tech -– expenditure approach} }
\vspace{-12mm}
\includegraphics[width=1\textwidth]{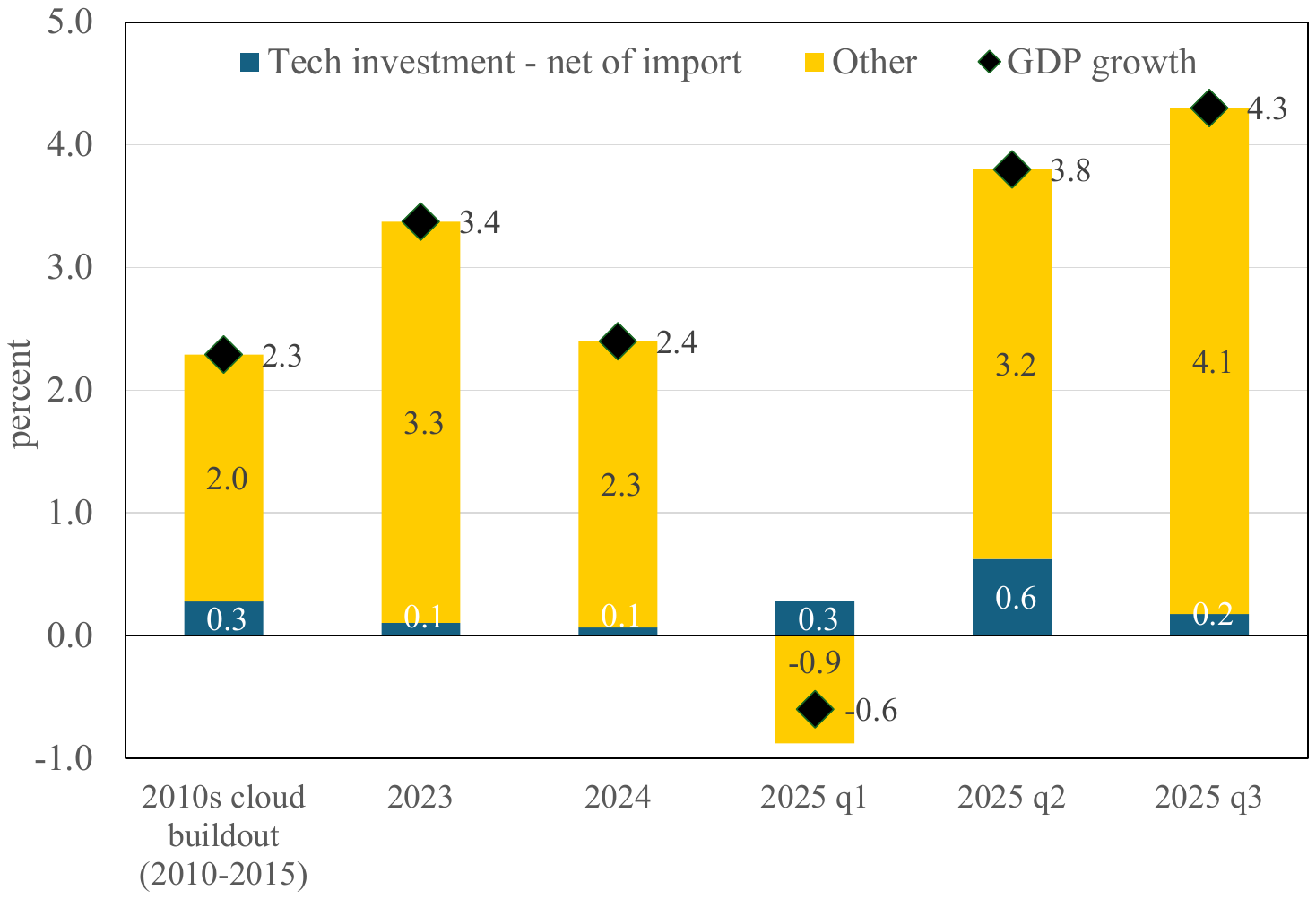}
\vspace{-12mm}
\raggedright
\footnotesize{\textit{\\Note: GDP growth rates are average quarterly annualized GDP growth over  different periods (2010q1--2015q4, 2023q1--2024q4, 2025q1, 2025q2, 2025q3). Contributions to GDP growth of tech and non-tech components (blue and orange bars), in percentage points, are averages of the contribution to quarterly annualized GDP growth in the same periods.}} 
\label{fig:gdpexp}
\end{center}
\end{figure}

Once tech investment is purged by tech imports, its contribution to GDP growth is downsized: it falls from above 1 percentage point to almost 0.3 in Q1, and more moderately in Q2 and Q3, to 0.6 and 0.2 percentage points, respectively. On average, its contribution has been around 20\% of overall GDP growth in the first 9 months of the year (Figure \ref{fig:gdpexp}). Calculations conducted by Evercore ISI for the first half of 2025 produce similar results.\footnote{They also assess that the contribution of the AI driven investment boom would have been 70\% without netting imports.} Although significantly rescaled, the contribution purged from imports is still spectacular, given that these investments represent about 4\% of overall GDP.
Furthermore, some measurement issues might lead to underestimate the contribution of these investments.\footnote{First, official statistics may understate AI investment if some key items, such as GPU servers, are misclassified as intermediate inputs rather than long‑lived capital goods. Statistical agencies might struggle to: (i) measure GPU servers’ price (due to rapid inflation of GPU costs), (ii) separate them from generic ICT imports – also due to purchases being based on confidential long‑term contracts, (iii) account for their massive price‑performance changes; (iv) record them in the detailed investment categories. Second, they may understate the US value added generated by AI‑related firms whose hardware production takes place abroad, if the profits and other income created in the United States are not fully recorded as exports of services.}
However, they have not taken the helm from the resilient American consumer, who added 0.4, 1.7, and 2.4 percentage points to GDP growth to the first three quarters of the year, respectively.

In the rest of the year and in 2026 capital expenditure is likely to keep sustaining production at least at the same pace: while US tech companies did not report specific targets for next year's capex, they all announced that they will be higher than in 2025.\footnote{Reporting q3 results, Alphabet, Amazon, Meta and Microsoft declared that their 2026 capex will be higher than 2025. For details, see \cite{IBD}.}

\section{From Data Centers to GDP: AI Production Channels}
\label{sec:production}

While Section~\ref{sec:investment} focused on the demand-side effects of AI-related capital expenditure and its import content, this section examines the broader question of how the ongoing production and use of AI services contribute to GDP. Beyond the build-up of data center capacity, the development and deployment of AI show up in national accounts when agents invest in AI-related research and development (R\&D), purchase AI-enabled services, and when these activities raise value added in technology-intensive sectors \citep{BEApaper}.

\begin{figure}[H]
\begin{center}
\caption{ \textsc{Flow of AI data center services to final use}}
\includegraphics[width=0.8\textwidth]{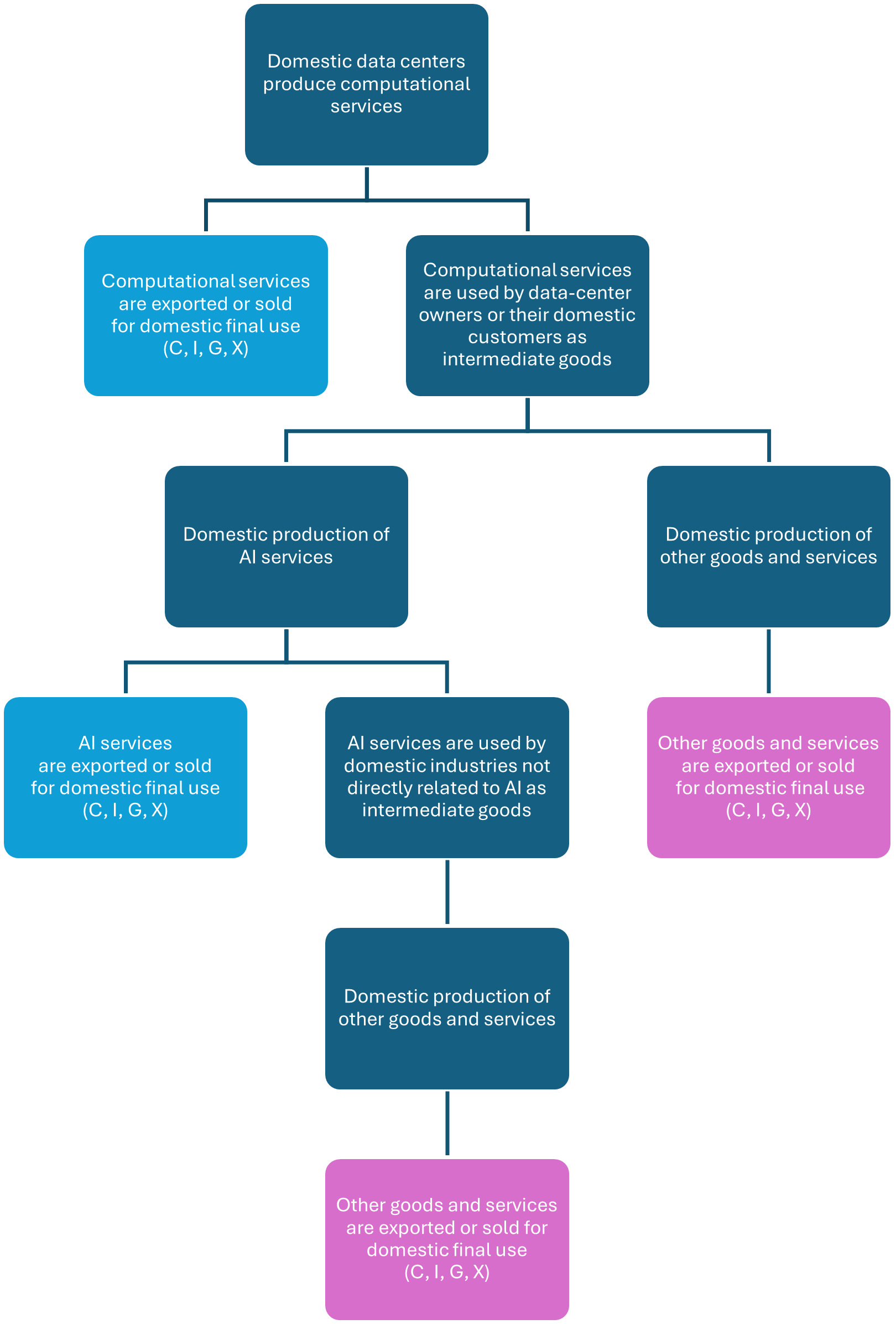}
\footnotesize{\textit{\\Note: Light blue boxes indicate cases where the increased production of computational and AI services provides direct positive contributions to GDP. Plum-colored boxes correspond to cases where the increased production is used as an intermediate input and does not provide direct contributions to GDP.}} 
\label{fig:flow}
\end{center}
\end{figure}

As explained previously, most of the AI services production activity involves data center utilization. For our purposes, the practical consequence is that this activity can be seen, also from an accounting perspective, as a downstream flow going from data centers, which supply computational capacity, all the way to AI services. The flow is illustrated in Figure \ref{fig:flow}.

There are two channels for computational capacity -- and the AI services that immediately consume that capacity -- to surface on GDP. 
First, in a very straightforward way, they can be sold for  final use (consumption, investment, exports or government expenditure), thereby contributing directly to GDP on the resource-use side (the light-blue boxes in Figure \ref{fig:flow}). A typical example is consumers purchasing ChatGPT subscriptions. Purchases of computational capacity for conducting R\&D -- provided they are recorded as such in corporate accounts -- also fall into this category, since R\&D is classified as investment in national accounts; further examples of such transactions are listed in Table \ref{tab:exa}. This is the channel that appears on GDP with the shortest lag relative to investment.\\

\begin{table}[h!]
\footnotesize
\centering
\caption{Examples of direct final demand contributions of data center and AI services}
\label{tab:direct-final-demand}
\begin{tabular}{p{4cm}p{5cm}p{5cm}} 
\toprule
\textbf{Contribution to final demand} & \textbf{National account categories} & \textbf{Examples (hypothetical)} \\
\midrule
Consumption & Personal consumption of other services (NIPA 2.4.5.95) & Consumer buys personal ChatGPT subscription \\ \\
Investment & Gross investment in intellectual property (NIPA 5.2.5U.19,47) & OpenAI trains a new model on Amazon’s cloud (R\&D expense) \\ \\
Government expenditure & Government consumption expenditures (NIPA 3.1.21) & Dept. of Defense runs its models on Microsoft’s cloud \\ \\
Exports & Exports of computer services (NIPA 4.2.5B.86) & A foreign firm uses Google’s cloud (on US-based data center) \\
\bottomrule \\
\end{tabular}
\label{tab:exa}
\end{table}

Second, computational capacity and AI services can be used as intermediate inputs by domestic producers of goods and non-AI services (the plum-coloured elements in Figure \ref{fig:flow}). For instance, a car manufacturer might rely on AI-driven production lines whose inference tasks are carried out in an AI data center, or a company may subscribe to ChatGPT for general employee use. In these cases, computational services do not appear directly in GDP as final resource use, although they may substantially alter firms' cost structures -- for example by reducing labor costs and increasing spending on intermediate goods. 

To the best of our knowledge, data on the precise distribution of computational services between final demand and intermediate use are currently unavailable. However, because the adoption of AI by non-AI-producing firms is still in its early stages, it is reasonable to assume that, at present, a large share of computational and AI services is sold directly to final users and thus rapidly surfaces in GDP statistics. 

To gauge the size of the potential contribution of AI services to final use, it is useful to consider two additional facts.  First, despite very large investments, all available evidence (e.g., \citealp{goldmansachs2025capacity}) points to high utilisation rates of new data centers and to persistent bottlenecks in meeting existing demand for computational capacity. The existing AI infrastructure is thus broadly at full capacity. Second, at current market prices for computational services (GPU rental), the  investment payback times at AI data centers is on the order of one year. Appendix \ref{app:econdatacenters} illustrates this with a simple calculation based on current cloud rental prices for NVIDIA's GB200 NVL72 architecture.\footnote{A modern high-density rack with 72 GPUs and 36 CPUs costs about USD~4.5~million to deploy once land, buildings, electrical systems, liquid cooling and other infrastructure are included, of which roughly two-thirds reflects the purchase of the server rack itself. Quoted annual rental prices for an equivalent rack from major cloud providers range from about USD~6.6~million to USD~10.1~million, while estimated yearly operating costs (electricity, maintenance, networking and other items) are on the order of USD~0.2~million. These figures are consistent with a payback period close to one year for fully utilised facilities.} 
Taken together, these two facts, along with the previous assumption that a large share of AI services is sold for final use, suggest that the yearly flow of revenues coming from the sale of AI services produced by these \textit{fully utilised} AI centers is of the same order of magnitude as capex. Therefore, in a matter of months it could produce remarkable GDP increases.

Despite the scarcity of disaggregated data on AI activity, some additional evidence helps to corroborate this picture. 
The presence of a ``revenue'' channel is supported by GDP data based on value added, in which the GDP for a given sector reflects not only investment in capital goods, but also the flow of goods and services generated by the existing capital stock. Although AI-related activities cannot be cleanly separated from the broader IT sector -- which likely leads to an overstatement of the contribution of the strictly defined AI industry -- the relevant aggregates point to a strong contribution from technology-intensive sectors. In particular, value added in the \textit{Manufacturing of computer and electronic products} , and in \textit{Computer systems design and related services} recorded a formidable increase in Q2, likely influenced by AI developments. Importantly, the value-added approach delivers a slightly higher contribution of tech to GDP than the expenditure approach: 0.5 percentage points in Q1 and 0.7 percentage points in Q2 (see Table \ref{tab:gdpva}), amounting on average to 40 per cent of GDP growth in the first half of the year (Figure \ref{fig:gdpva}),\footnote{At the time of writing, the Bureau of Economic Analysis had not yet released the data of Gross Domestic Product by Industry for Q3.} notably more than the figure recovered through the expenditure approach that only captures investment.

\begin{table}[H]
\centering
\caption{Contributions to GDP growth by sector.}
\label{tab:gdpva}
\footnotesize
\scalebox{.8}{%
\begin{tabular}{lcccc}
\toprule
 & \multicolumn{2}{c}{2025 Q1} & \multicolumn{2}{c}{2025 Q2} \\
\cmidrule(r){2-3} \cmidrule(r){4-5}
 & q/q growth ($\%$) & contr. (p.p.) & q/q growth ($\%$) & contr. (p.p.)\\
\midrule
GDP & -0.6 & & 3.8 & \\
\addlinespace
\textbf{Contribution to GDP growth:} & & & & \\
\quad Private Industries & & & & \\
\quad\quad Manufacturing & & & & \\
\quad\quad\quad \textit{Computer and electronic products} & 3.5 & 0.03 & 17.2 & 0.17 \\
\addlinespace
\quad\quad Information & & & & \\
\quad\quad\quad \textit{Data proc., internet pub., other info. services} & 14.1 & 0.24 & 7.7 & 0.14 \\
\addlinespace
\quad\quad Professional and business services & & & & \\
\quad\quad\quad \textit{Computer systems design and related services} & 12.2 & 0.21 & 23.4 & 0.38 \\
\addlinespace
\midrule
Total contribution & & 0.48 & & 0.69 \\
\bottomrule
\end{tabular}}
\raggedright
\footnotesize{\textit{\\\vspace{0.5cm} Note: Quarter-on-quarter growth rates are annualized and computed on real value added values. Contributions to quarterly annualized investment growth are expressed in percentage points. ``Data proc., internet pub., other info. services'' stands for ``Data processing, internet publishing, and other information services''.}} 

\end{table}

\begin{figure}[H]
\begin{center}
\caption{ \textsc{Quarterly data center revenues of the 3 largest public clouds}}
\vspace{-12mm}
\includegraphics[width=0.9\textwidth]{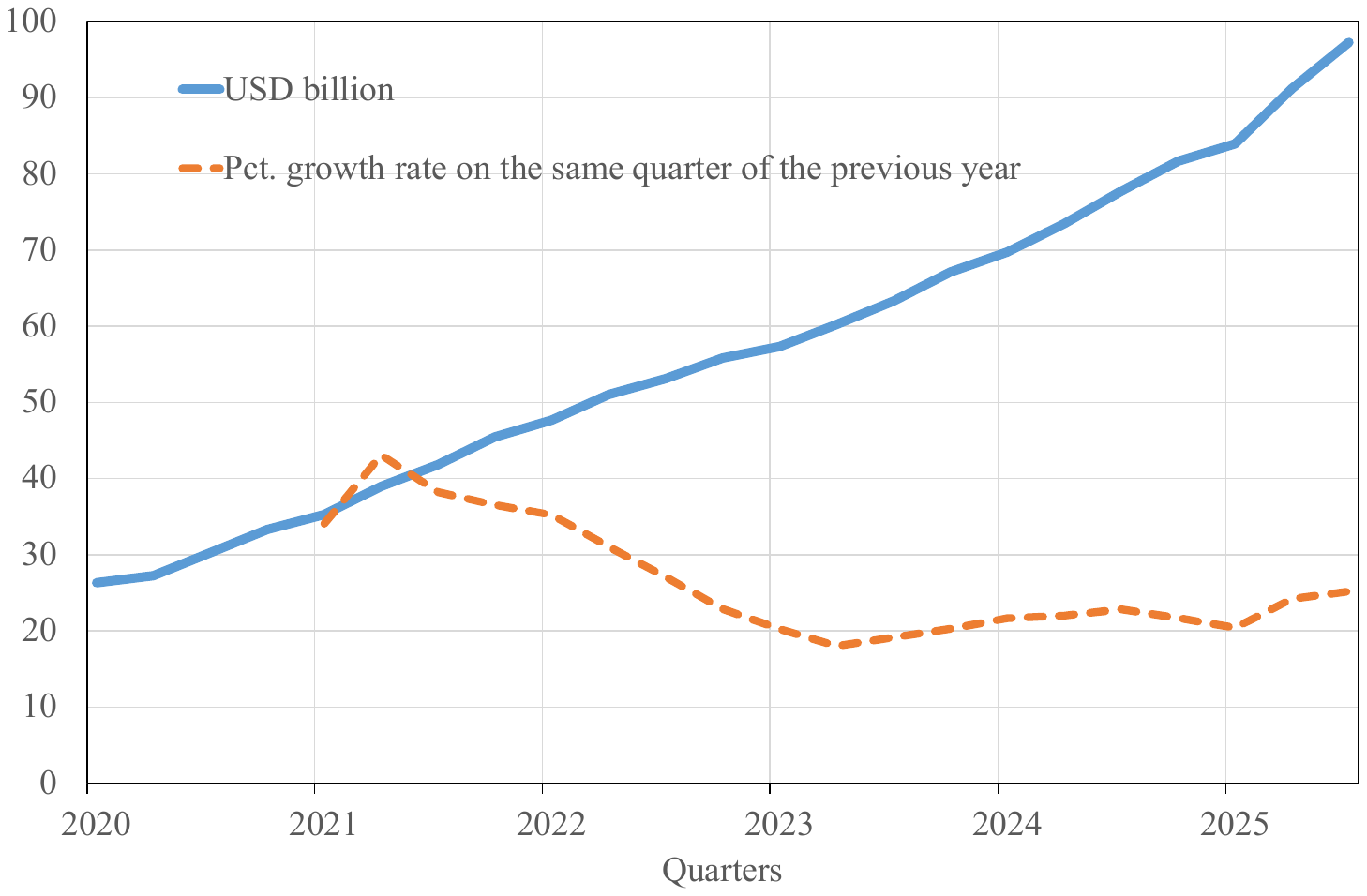}
\vspace{-15mm}
\raggedright
\footnotesize{\textit{\\Note: The three largest public clouds are Amazon AWS, Google Cloud and Microsoft Cloud. Revenues generated by these clouds are reported as a separate item in their respective owner companies’ financial statements.  Source: quarterly financial reports of Amazon, Alphabet (Google), Microsoft and own calculations.}} 
\label{fig:topthree}
\end{center}
\end{figure}

Moreover, the revenues generated by the three largest public clouds that rent AI servers continue to grow at rates well above 20~per cent, despite already reaching impressive levels (around USD~350~billion over the past four quarters; Figure \ref{fig:topthree}). These revenues represent a large share of the public cloud market (about 65~per cent according to \citealp{omdia2025cloudspending}), but by no means the majority of the total data center market, which also includes large private clouds (e.g.\ those operated by Apple, Meta and Tesla) and smaller data centers owned by other corporations. Many of these private clouds are used to produce AI services sold to final users. Together with smaller public clouds, they likely make the aggregate volume of computational and AI services that may affect GDP much larger than the USD~350~billion figure cited above.

\begin{figure}[H]
\begin{center}
\caption{ \textsc{Exports of computer services}}
\vspace{-12mm}
\includegraphics[width=1\textwidth]{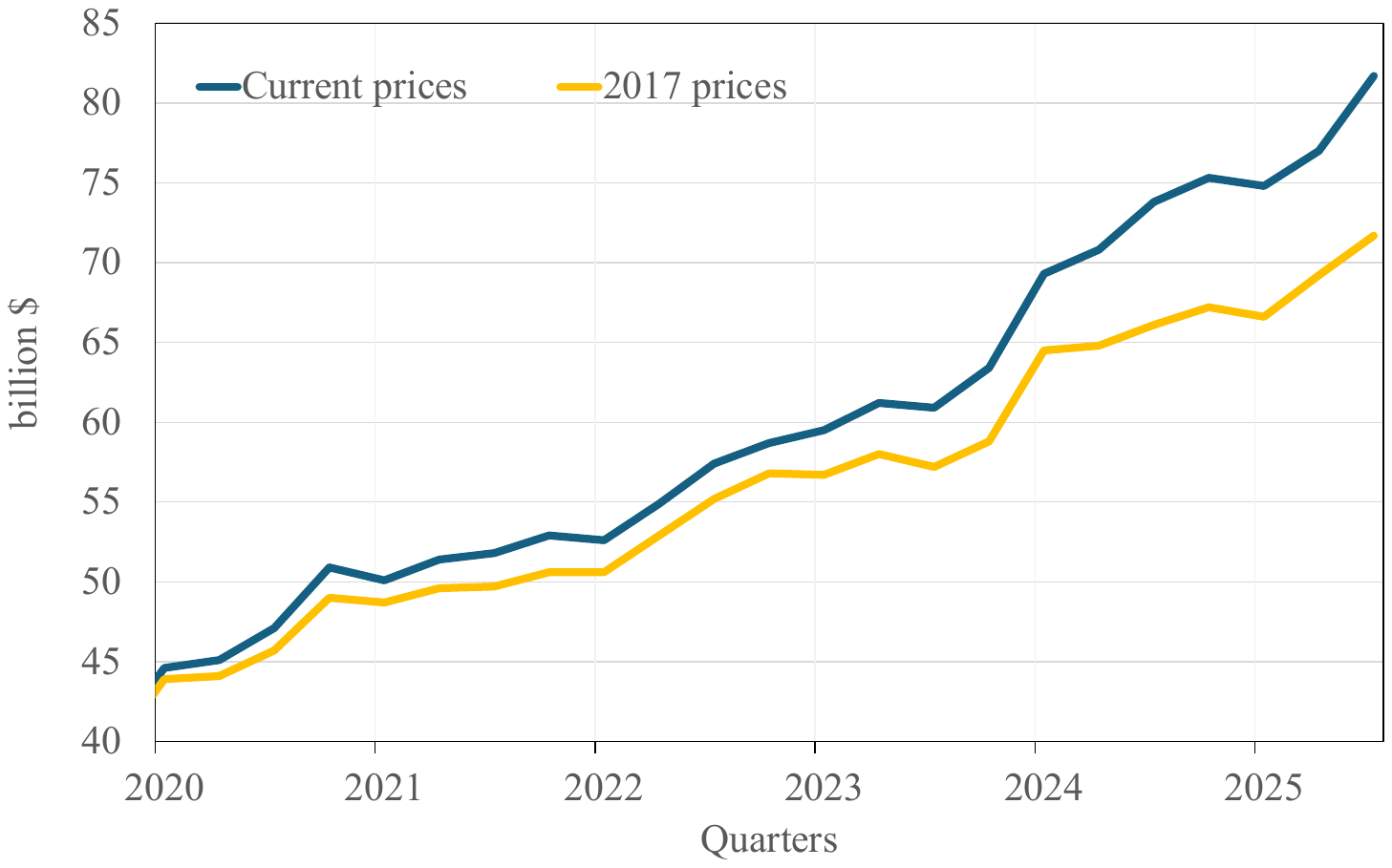}
\vspace{-15mm}
\label{fig:expcompserv}
\end{center}
\end{figure}

Finally, national accounts aggregates that provide more granularity also point to growing export flows of AI-related services. In particular, the ``Exports of computer services'' aggregate, which includes cloud services, reached an annualized\footnote{Annualized rates reported by the Bureau of Economic Analysis roughly correspond to quarterly amounts multiplied by 4.} rate of USD~82~billion in the third quarter of 2025, up 11~per cent from the same quarter of the previous year (Figure \ref{fig:expcompserv}). Although these data again bundle AI with broader digital services, they are consistent with the view that the production of computational and AI services is becoming an increasingly important driver of US domestic and external demand.

\section{What to expect going forward}
\label{sec:future}

As discussed above, in national accounts, AI shows up first not as a productivity boost but as spending. It works mainly through expenditure-side channels: higher investment in software, data, and AI-related capital, increased consumption of AI-enabled services. Deeper effects connected to productivity gains and resource reallocation are slower to materialize, as organizations adapt, processes are redesigned, costs fall persistently, new products are enabled, and demand adjusts. It is these second-round changes—rather than the initial wave of AI investment—that will ultimately determine whether AI leaves a durable imprint on growth. This is the big existential question on AI and the macroeconomy.

However, also the current phase of rising demand, with its staggering spending volumes, poses some very pressing macroeconomic questions.  
A first issue revolves around the depreciation of the AI hardware: servers and other computing equipment may have relatively short economic lifetimes because the continuous arrival of more powerful and efficient hardware makes existing capital obsolete fairly quickly.\footnote{Although replacement cycles may nowadays be longer due to the slowing down of Moore's law.}  Technological obsolescence may be compounded by accelerated physical wear, driven by extreme thermal stress that would increase GPUs failure rates. Investors and commentators have pointed to these short lifetimes as a potential source of fragility. It would increase the risk that capital cannot be fully amortized through operating cash flows before the next technology cycle, if the revenue growth fails to translate into adequate returns on invested capital (ROIC).\footnote{See for example this discussion by  \citet{Krugman2025} and this Bloomberg interview to \citet{Chanos}} If insufficiently reflected in balance sheets, this dynamic could temporarily inflate reported profits, sustain excessive valuations, and heighten the risk of abrupt corrections once the effective depreciation is recognized.\footnote{See this WSJ article by \citet{WSJ2025}.} Moreover, frequent replacement would mechanically raise capex-to-revenue ratios and depress free cash flow, limiting the share of operating profits that can ultimately be distributed to investors—even in high-growth environments.

Some of these concerns may be overstated. As documented above, short economic life cycles are usually offset by equally short payback periods: the revenues and profits generated by a new data center typically exceed the initial outlay after only a few quarters, a pattern that appears to hold also for the latest generation of facilities, as discussed in the Appendix. In addition, GPUs may retain economic value well beyond their technological frontier, as older chips can be redeployed across less demanding tasks over time, effectively extending their amortization horizon. Taken together, these factors substantially mitigate concerns related to ROIC and accounting depreciation. 

By contrast, the nexus between accelerated reinvestment cycles and free cash flow appears more robust. Although data center capex and the associated revenues have large effects on gross income, their impact on net income is likely to be much smaller when viewed over a sufficiently long horizon, given that investment costs must be incurred with a higher frequency than in more traditional investment frameworks. Furthermore, when data center investments are ramped up in response to a permanent increase in demand for computational services, this should not be seen as a temporary spike: it entails a sustained and rapid pace of replacement investment over time. It should also be noted that, to the extent that hardware production remains highly fragmented along globalized value chains, a substantial share of these large investment waves is likely to continue accruing to foreign suppliers and partners, rather than translating into commensurate gains for the domestic economy. As a result, the macroeconomic spillovers of the AI-related investment would continue to be more limited than suggested by headline capex figures for income generation at home.

The second critical issue revolves around the pace of development of AI services’ demand. A distinctive feature of the current AI revolution, compared with earlier technological transformations such as the diffusion of personal computers in the 1980s and the Internet in the 1990s, is the extraordinary speed of adoption. As of end 2025, only three years have passed since the launch of ChatGPT (at the end of 2022), which marked the first major deployment of advanced generative AI services to a mass audience. Yet, according to most estimates (e.g., \citealp{microsoft2025aidiffusion}), the global user base of these services has already surpassed one billion people. By contrast, following the birth of the World Wide Web and the release of the first web browsers in the early 1990s, it took roughly 15 years for Internet services to reach a comparable level of penetration.

This unprecedented pace of adoption has created significant challenges for AI providers, which must satisfy booming demand while securing sufficient computational resources. The wave of large scale AI investments observed in 2025, particularly in data center capacity, can be interpreted as a direct response to these pressures.

The outlook of future demand is therefore the key variable for AI firms' decisions. Yet it is surrounded by an extremely high degree of uncertainty. As Anthropic’s CEO Dario Amodei noted in an interview with the New York Times \citep{nyt2025anthropic}, when demand for a service expands by around 1,000\% annually for two consecutive years (in 2024 and 2025), it becomes nearly impossible to forecast the following year’s (i.e., 2026) growth rate with economically meaningful precision. AI firms are therefore compelled to design investment and procurement plans under uncertainty. Risks are two-sided. On the one end of the spectrum, there is the much debated risk of overinvestment: if AI capacity were to exceed actual demand, the economics of AI production could deteriorate, potentially triggering market corrections and episodes of financial stress. In this gloomy scenario, emphasized by several commentators,\footnote{See, e.g., \citet{brooker2025aibubble}, \citet{laudani2025aibubble}, \citet{letzing2025aibubble}.} smaller and less established players, such as independent AI labs, could be more vulnerable, while larger conglomerates that can count on diversified revenue streams would likely be better positioned to absorb the shock, although their margins, which are currently supported by scarcity and buoyant demand, could shrink significantly. Among the large players, some that are almost exclusively focused on AI chip production could see their profits contract more rapidly. Repercussions on stock prices could be material and potentially induce second-order macroeconomic effects. 

On the other end, there is the --  often overlooked -- opposite scenario of underinvestment. This configuration also presents substantial danger for AI firms. 
If planned computational resources ultimately were to fall short of a particularly strong expansion in demand, the resulting flow of AI services could be of lower quality\footnote{E.g., user queries could be routed to weaker models.} and potentially unreliable\footnote{E.g., API requests could be throttled during peak hours.}. Alternatively, there could be upward adjustments in selling prices. Either of these outcomes would likely drive users away and firms could risk ceding ground to competitors, including Chinese firms.

Which of these risk scenarios will materialize is difficult to assess, not least because limited transparency and the scarcity of detailed data on the evolution of demand for AI services constrain independent analysis and forecasting. Economic intuition suggests that the rapid, ongoing improvement in AI capabilities,\footnote{\citet{kwa2025measuring}, for example, show that a measure of the complexity of tasks that AI systems can tackle with satisfactory accuracy doubles every 7 months. Other metrics show similar dynamics. See, e.g., the model scores regularly published by the independent AI-benchmarking company Artificial Analysis on their web site artificialanalysis.ai.} combined with the fact that enterprise adoption is still far from complete, could continue to sustain robust growth in demand. If so, the macroeconomic implications of AI will become even more significant than they are today and will warrant close scrutiny by policymakers, including dedicated efforts to collect more granular data on the AI industry.

\section{Conclusions and policy implications}
\label{sec:conclusions}

This paper has provided a first pass at mapping the macroeconomic footprint of the current AI wave in the United States.
Bringing together a mechanical macro-accounting framework and a stylised description of the AI production ecosystem, we have traced how AI-related activity already appears in US GDP through investment in digital infrastructure and the provision of computational and AI services.

Three broad messages emerge. 
First, AI investment has played a major role in sustaining demand and activity in the first three quarters of 2025, but it has not been the sole engine of US growth.  However, once the high import content of AI-related hardware is taken into account, the contribution of technology-related investment to GDP growth, while still substantial, is more limited than suggested by headline spending and by some popular narratives. 

Second, data centers are the backbone of the AI ecosystem. They should therefore be central to any macroeconomic analysis of AI’s effects. On the one hand, the construction and equipment of AI data centers provide a considerable stimulus to aggregate demand and GDP through the investment channel. On the other hand, data centers enable a flow of computational and software services that almost immediately and significantly feed into final consumption and other components of aggregate demand. 

Third, given high utilization rates and current pricing of GPU-based services, simple calculations suggest that the revenues generated by these facilities could soon support GDP increases of similar magnitude to the underlying capital expenditures. Sectoral value-added data and trade in computer services already point in this direction.

These findings have several implications for policymakers. For monetary policy, they caution against attributing US economic resilience solely to AI, as the resilient US consumer has remained a central driver of the economic dynamics. At the same time, they show that AI-related investment and production can meaningfully affect both demand and supply conditions. For fiscal and industrial policy, they highlight the importance of the location of high-value segments of the AI value chain, and the potential role of infrastructure, skills and regulatory frameworks in affecting location choices.

The analysis has clear limitations. It abstracts from longer‑term productivity effects, from labor‑market adjustments, and from distributional consequences of AI adoption; furthermore, it relies on sectoral aggregates that imperfectly isolate AI from the broader technology sector, and on partial information on firms' investment and revenue plans.  
Addressing these gaps will require better statistics -- including improved classification of AI‑related hardware and services in national accounts, and greater use of micro‑level data -- as well as richer modelling of AI diffusion and its interactions with macroeconomic dynamics.  

Nevertheless, the perspective developed here can serve as a framework for interpreting incoming data and for structuring future analysis.  
As AI technologies continue to evolve and diffuse, a clear understanding of how they map into the standard macroeconomic accounts will be essential for timely and well-informed policy decisions.

\clearpage

\typeout{}
\bibliographystyle{apalike}
\bibliography{CNTAIbib.bib}
\clearpage

\appendix

\section{The economics of modern AI data centers} \label{app:econdatacenters}

\counterwithin{figure}{section}
\renewcommand{\thefigure}{A\arabic{figure}}
\renewcommand{\thetable}{A\arabic{table}}
\setcounter{figure}{0}
\setcounter{table}{0}


A typical last-generation AI data center is built around NVIDIA’s NVL72 architecture: high-density pre-assembled server racks (i.e., small shelves packed with computer equipment) that contain 72 GPUs and 36 CPUs. Each of these racks occupies less than 1 square meter of floor space and costs about 3 USD million. The purchase price of these NVIDIA’s racks represents the majority of the total capital expenditure needed to build a modern AI data center. Table~\ref{tab:capex-per-rack} provides an estimate of the break-down of the total capital expenditure needed to build an AI data center, per server-rack. Table~\ref{tab:rental-prices} reports the rental prices of server racks, as quoted by some cloud infrastructure providers. Table~\ref{tab:opex-per-rack} reports estimates of the yearly operating costs of a rack. Putting together these estimates, we can infer that, at current on-demand prices, the estimated payback period of an AI data-center investment is likely to be around 1 year, with a considerable margin for error due to several factors that are difficult to measure but should not radically change the estimate.\footnote{To be precise, the average yearly rent of an NVL72 rack at current on-demand prices largely exceeds the sum of the capital expenditure and the yearly operating cost. To infer the investment payback period precisely, the following information should also be known: 1) organizational overheads and the cost of developing and maintaining the cloud-management software used to rent the instances (the incidence of these costs decreases as scale increases); 2) the proportion of idle time (which should be small given reports of still binding capacity constraints); 3) the proportion of GPUs that are unavailable due to failures (although there are usually warranties in place, and NVL72 systems have sophisticated mechanisms to bypass failed components); 4) the proportion of capacity rented through long-term contracts and the average discounts applied to these contracts; 5) the additional revenue generated by network, storage and orchestration charges; these revenues are not included in the headline hourly costs reported here because they are highly user- and workload-specific and because they may be partly produced by separate hardware whose costs are not accounted for in Table A1.}

\begin{table}[h!]
    \centering
    \caption{Data-center capital expenditure (per NVL72 server rack)}
    \label{tab:capex-per-rack}
    
    \begin{tabular}{lrr}
        \hline
        & \multicolumn{1}{c}{USD million per rack} & \multicolumn{1}{c}{Per cent} \\
        \hline
        NVIDIA GB200 NVL72 server rack & 3.00 & 67\% \\
        Land                            & 0.12 & 3\%  \\
        Building                        & 0.24 & 5\%  \\
        Electrical systems              & 0.53 & 12\% \\
        Liquid cooling and other HVAC   & 0.38 & 8\%  \\
        Other                           & 0.24 & 5\%  \\
        \hline
        Total                           & 4.50 & 100\% \\
        \hline
    \end{tabular}

    \vspace{0.5em}
    \begin{minipage}{0.9\textwidth}
        \footnotesize
        \textbf{Source}: Savills, Cushman \& Wakefield, Dgtl Infra, Data Centre Dynamics, Anastasi In Tech, own calculations.\\
        \textbf{Note}: estimated costs are obtained by averaging estimates from various sources and applying correction factors for liquid-cooled high-density equipment when estimates refer to traditional data centers. These figures refer to US-based data centers and do not include capital expenditures for dedicated power generation plants (i.e., power is assumed to be bought from third parties).
    \end{minipage}
\end{table}

\begin{table}[h!]
    \centering
    \small
    \caption{Rental prices (USD) of NVL72 server racks}
    
    \label{tab:rental-prices}
    \begin{tabular}{lrrr}
        \hline
        Cloud provider & Price per hour & Price per day & Price per year \\
        \hline
        EC2       & 762  & 18,288 & 6,675,120 \\
        CoreWeave & 756  & 18,144 & 6,622,560 \\
        Oracle    & 1,152 & 27,648 & 10,091,520 \\
        \hline
    \end{tabular}
    
    \vspace{0.5em}
    \begin{minipage}{0.9\textwidth}
           \centering 
        \footnotesize
        \textbf{Source}: web sites of cloud infrastructure providers.\\
        \textbf{Note}: prices as quoted by cloud providers in October 2025.
    \end{minipage}
\end{table}

\begin{table}[h!]
    \centering
    \small
    \caption{Estimated yearly operating cost of an NVL72 server rack}
    \label{tab:opex-per-rack}
    
    \begin{tabular}{lr}
        \hline
        Cost component & Yearly cost (USD) \\
        \hline
        Electricity & 105,000 \\
        Maintenance & 25,000  \\
        Network     & 20,000  \\
        Other       & 40,000  \\
        \hline
        Total       & 190,000 \\
        \hline
    \end{tabular}
        
    \vspace{.5cm}
    \begin{minipage}{0.9\textwidth}
       \centering 
        \footnotesize
        \textbf{Source}: authors' estimates based on multiple sources.
    \end{minipage}
\end{table}

\section{Value added approach} \label{app:vaapproach}

\begin{figure}[H]
\begin{center}
\caption{\textsc{Contribution to growth, tech vs non-tech -– value added approach} }
\label{fig:gdpva}
\vspace{-15mm}
\includegraphics[width=1\textwidth]{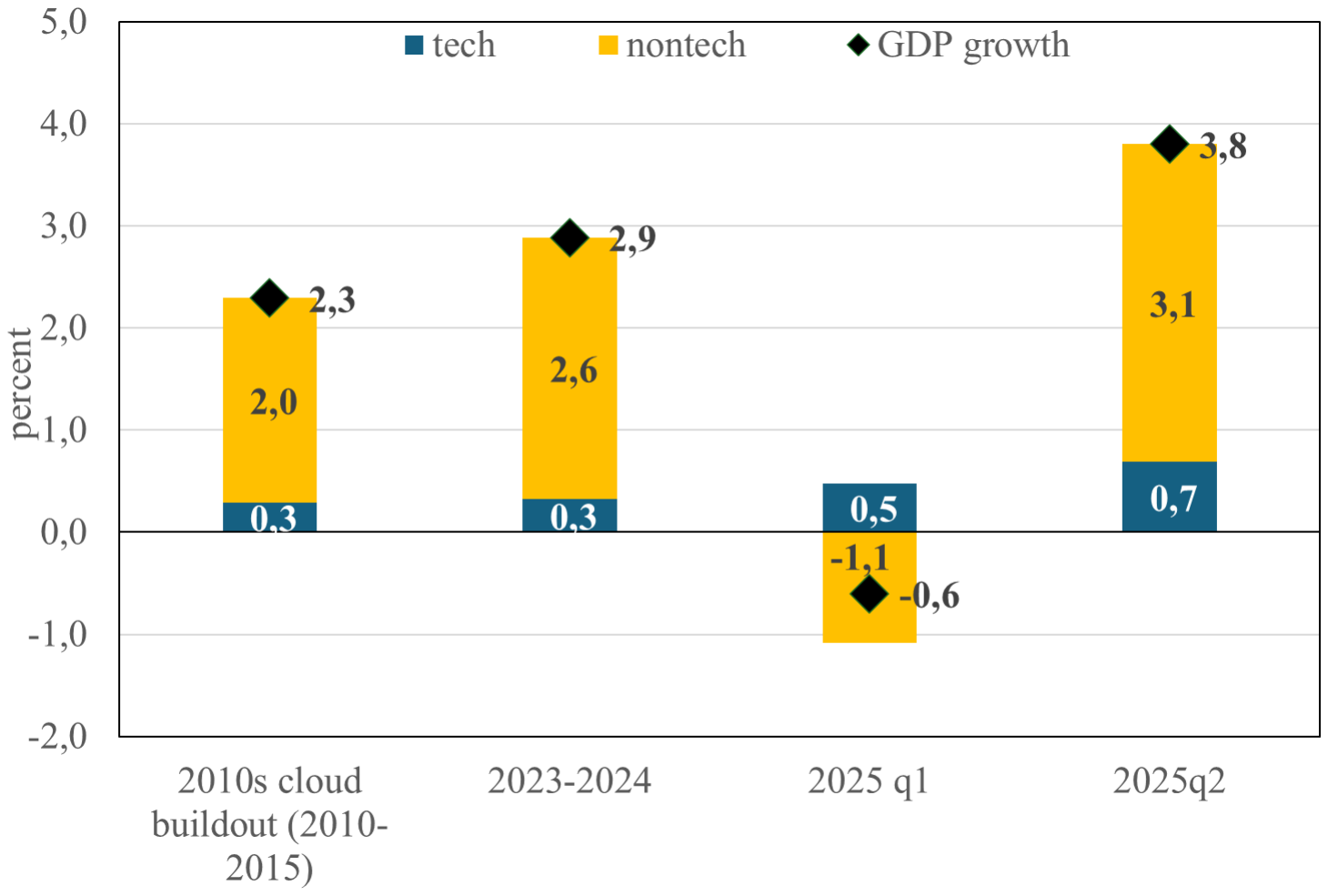}
\end{center}
\vspace{-15mm}
\footnotesize{\textit{\\Note: GDP growth rates are average quarterly annualized GDP growth over three different periods (2010q1--2015q4, 2023q1--2024q4, 2025q1--2025q2). Contributions to GDP growth of tech and non-tech components (blue and orange bars), in percentage points, are averages of the contribution to quarterly annualized GDP growth in the same periods.}} 

\end{figure}

\end{document}